\def\e{{\rm e}}
\def\del{\partial}
\def\half{{1\over2}}

\def\vev#1{\langle #1 \rangle}
\def\del{\partial}
\def\half{{1\over2}}

\def\vev#1{\langle #1 \rangle}

\def\del{\partial}
\def\dslash{\del\kern-0.55em\raise 0.14ex\hbox{/}}

\def\rough#1{\raise.3ex\hbox{$#1$\kern-.75em\lower1ex\hbox{$\sim$}}}

\newcommand{\PRD}[3]{{\it Phys. Rev.} {\bf D{#1}} (19{#3}) {#2}}

\newcommand{\PRDM}[3]{{\it Phys. Rev.} {\bf D{#1}} {#2} (20{#3})}

\newcommand{\PRLM}[3]{{\it Phys. Rev. Lett.} {\bf {#1}} {#2} (20{#3})}
\newcommand{\NPB}[3]{{\it Nucl. Phys.} {\bf B{#1}} {#2} (19{#3})}
\newcommand{\NPBM}[3]{{\it Nucl. Phys.} {\bf B{#1}} (20{#2}) {#3}}
\newcommand{\PLB}[3]{{\it Phys. Lett.} {\bf B{#1}} (19{#2}) {#3}}
\newcommand{\PLBM}[3]{{\it Phys. Lett.} {\bf B{#1}} (20{#2}) {#3}}

\newcommand{\PTP}[3]{{\it Prog. Theor. Phys.} {\bf {#1}} (19{#3}) {#2}}
\newcommand{\PTPM}[3]{{\it Prog. Theor. Phys.} {\bf {#1}} (20{#3}) {#2}}
\newcommand{\ANN}[3]{{\it Ann. Phys. (N.Y.)} {\bf {#1}}, {#2} (19{#3})}

\newcommand{\MPL}[3]{{\it Mod. Phys. Lett.} {\bf A{#1}} (19{#3}) {#2}}

\newcommand{\jhep}[3]{{\it JHEP} {\bf {#1}} (20{#2}) {#3}}

\newcommand{\hepph}[1]{{\tt hep-ph/#1}}

\newcommand{\hmu}{\hat\mu}

\documentclass[12pt]{article}
\usepackage{graphicx}
\begin{document}
\baselineskip=18pt
\begin{titlepage}
\begin{flushright}
KOBE-TH-07-03\\
TU-793
\end{flushright}
\vspace{1cm}
\begin{center}{\Large\bf 
On Gauge Symmetry Breaking via Euclidean Time Component
of Gauge Fields
}
\end{center}
\vspace{1cm}
\begin{center}
Makoto Sakamoto$^{(a)}$
\footnote{E-mail: dragon@kobe-u.ac.jp} and
Kazunori Takenaga$^{(b)}$
\footnote{E-mail: takenaga@tuhep.phys.tohoku.ac.jp}
\end{center}
\vspace{0.2cm}
\begin{center}
${}^{(a)}$ {\it Department of Physics, Kobe University, 
Rokkodai, Nada, Kobe 657-8501, Japan}
\\[0.2cm]
${}^{(b)}$ {\it Department of Physics, Tohoku University, 
Sendai 980-8578, Japan}
\end{center}
\vspace{1cm}
\begin{abstract}
We study gauge theories with/without an extra dimension
at finite temperature, in which there are two kinds of order
parameters of gauge symmetry breaking. One is the zero mode
of the gauge field for the Euclidean time direction, and the other
is that for the direction of the extra dimension. We evaluate the
effective potential for the zero modes in the one-loop approximation 
and investigate the vacuum configuration in detail. Our analyses
show that gauge symmetry can be broken only through the zero mode
for the direction of the extra dimension, and no nontrivial vacuum
configuration of the zero mode for the Euclidean time direction
is found.
\end{abstract}
\end{titlepage}
\newpage
\section{Introduction}
When one considers quantum field theory on space-time where some of
spatial coordinates are compactified on certain topological manifolds, one
expects rich dynamical phenomena which shed a new insight and
give a deep
understanding on the fundamental, long-standing problems in 
high energy physics. In fact, it has been shown that a new mechanism of 
spontaneous supersymmetry breaking can occur \cite{sakamoto1}, and 
various phase structures arise in field theoretical models on such 
space-time \cite{sakamoto2, sakamoto3}. Hence, that is an 
important research theme.
\par
One of the famous dynamical phenomena is the Hosotani
mechanism \cite{hosotani}. Zero modes of component gauge
fields for compactified directions become dynamical 
variables due to topology of extra dimensions, so that 
they cannot be gauged away. A remarkable physical 
consequence of it is that the zero modes, which
are closely related to the Wilson line phases, can develop a 
vacuum expectation value (VEV), so that gauge symmetry can be 
broken through the VEV.
\par
The VEV is determined by the minimum of the effective potential.
The gauge symmetry 
breaking patterns induced by the VEV have been studied extensively 
in (supersymmetric) gauge theories \cite{gaugesym1, gaugesym2}, and they 
depend on matter content in the theory and boundary 
conditions of fields for compactified directions. 
Recently, much attention has been paid to the idea
of the gauge-Higgs unification based on the Hosotani mechanism, where
the Higgs field is unified into part of higher dimensional gauge 
fields, as a desirable candidate of physics
beyond the standard model \cite{gh1, gh2, gh3}. The 
gauge-Higgs unification can (re)solve shortcomings 
in the Higgs sector of the standard model.   
\par
It is interesting to investigate higher dimensional gauge
theories in the context of, for instance, the early universe 
by incorporating the temperature into the framework.
Then, we may have another source of the gauge symmetry breaking.
In finite temperature field theory, the Minkowski time $t$
is replaced by the Euclidean time $-i\tau$, and it is compactified
on a circle $S^{1}_{\tau}$ whose length of the circumference is the
inverse temperature $T^{-1}$.
If one considers gauge theories at finite temperature, one
immediately realizes that the zero mode of the component gauge
field $A_{\tau}$ for the $S^{1}_{\tau}$ (Euclidean time) direction 
becomes a dynamical variable due to the topology of the circle and
can be a source of the gauge symmetry breaking.
\par
Finite temperature field theory has a resemblance to quantum field 
theory on topological spatial extra dimensions in the sense that the
zero mode of the component gauge field $A_{\tau}$ 
for the compactified Euclidean time direction is a dynamical variable.
There is, however, a crucial difference between the two theories. In 
finite temperature field theory, the boundary conditions of the fields for 
the $S_{\tau}^1$ (Euclidean time) direction are uniquely fixed by the quantum 
statistics; that is, one must assign (anti)periodic boundary conditions 
for (fermions) bosons \cite{finitet, kugo},
while one does not know, {\it a priori}, the boundary conditions 
for the compactified spatial direction.
As we mentioned above, since 
gauge symmetry breaking can occur through the VEV of the
gauge field for the compactified spatial direction,
we are very much interested in 
the possibility of gauge symmetry breaking
through the VEV of the gauge field $A_{\tau}$ for the Euclidean time 
direction.
\par
The one-loop effective potential for $\vev{A_{\tau}}$ has been 
derived in the $SU(N)$ gauge theory at finite temperature
with/without an extra 
dimension \cite{gross,weiss,altes,fara}. In \cite{gross}, the
color screening effects are studied in the four dimensional
$SU(N)$ gauge theory with the adjoint and fundamental 
fermions at finite temperature. Refs. \cite{weiss,altes,fara}
have paid attention to the global $Z_N$ symmetry whose breaking 
is a signal of the deconfinement phase. The models considered 
there are restricted to the 
$SU(N)$ gauge theories that possess the $Z_N$ symmetry, which
is the center of $SU(N)$.
\par
We are interested in $\vev{A_{\tau}}$ {\it as the source of 
gauge symmetry breaking}, unlike the 
references \cite{gross,weiss,altes,fara}. The purpose of this paper 
is to investigate the possibility of the gauge symmetry breaking 
via nontrivial VEVs of the component gauge field $A_{\tau}$ 
for the Euclidean time direction in finite temperature gauge
theories with/without an extra dimension. It has been known
that gauge symmetry breaking can occur through nontrivial
VEVs of the component gauge fields for the compactified 
spatial direction \cite{hosotani,quiros}. It is interesting 
to examine whether or not the extra dimensions affect the VEV of the 
component gauge fields $A_{\tau}$ for the Euclidean time
direction.
\par
In section $2$, we explain that the zero mode of $A_{\tau}$ is 
actually a dynamical variable. In section 3, we evaluate 
the effective potential for the zero mode of $A_{\tau}$ in 
perturbation theory and study the gauge symmetry breaking 
patterns. The models we study are the $D$ dimensional $SU(N)$ 
gauge theories at finite temperature with massless/massive 
bosons and fermions belonging to the adjoint and fundamental 
representation and the $SU(2)$ gauge theory with the massless 
fermion belonging to the higher dimensional representation 
under the $SU(2)$ at finite temperature. We find no nontrivial 
VEV is obtained for the models. In section $4$, we consider 
one extra dimension at finite temperature in order to 
investigate the possibility of having a  nontrivial VEV for the 
zero mode for the Euclidean time direction, and numerically 
study the gauge symmetry breaking in the $SU(2)$ gauge theory with 
massless/massive adjoint and fundamental fermions. The final 
section is devoted to conclusions.
\section{$A_{\tau}$ as a dynamical variable at finite temperature}
In this section, we briefly review the discussion that 
the zero mode of the component gauge field for the Euclidean 
time direction is a dynamical variable 
%
%
and present one of the the physical consequences of nontrivial 
values of the zero mode.
To this end, we consider an $SU(N)$ gauge theory on $D$
dimensions at finite 
temperature. Then, the Euclidean time direction is compactified on 
$S^1$, which is hereafter denoted by $S^1_{\tau}$, so that we 
study the gauge theory on $S_{\tau}^1\times R^{D-1}$, where $R^{D-1}$ 
is the $D-1$ dimensional flat space. The length of the 
circumference of the $S_{\tau}^1$ is given by the inverse temperature 
$\beta\equiv T^{-1}$. The boundary condition of the
gauge field $A_{\hmu}$ must satisfy the periodic boundary condition,
\begin{equation}
A_{\hmu}(x^i, \tau +\beta)=A_{\hmu}(x^i,\tau),
\label{shiki1}
\end{equation}
where ${\hmu}~(i)$ runs from $1$ to $D~(D-1)$. 
The gauge field $A_{\hmu}$ is decomposed 
as $(A_i,~A_D\equiv A_{\tau})~(i=1,\cdots, D-1)$ \footnote{The component
gauge field $A_{\tau}$ is related with the time component $A_0$ in the 
Minkowski space-time by $A_0=iA_{\tau}$.}. $A_{\tau}$ is the
component gauge field for the Euclidean time direction.
\par
Let us consider the gauge transformation,
\begin{equation}
A_{\hmu}\rightarrow A_{\hmu}^{\prime}
=U\left(A_{\hmu}+{i\over g}\del_{\hmu}\right)U^{\dagger},
\label{shiki2}
\end{equation}
where $g$ is the $D$ dimensional gauge coupling. Here we require 
that the gauge transformation function $U$ must satisfy the 
periodic boundary condition 
\begin{equation}
U(x^i, \tau+\beta)=U(x^i,\tau);
\label{shiki3} 
\end{equation}
Otherwise, the gauge transformation would change the boundary
condition of the fields. Reflecting the topology 
of $S_{\tau}^1$, we can also consider the 
following gauge transformation function,
\begin{equation}
U(x^i, \tau)=
{\rm diag}.\left(\e^{i 2\pi T m_1\tau}, \e^{i 2\pi T m_2\tau},
\cdots, \e^{i 2\pi T m_N\tau}\right) \quad \mbox{with}\quad 
\sum_{i=1}^N m_i=0,
\label{shiki4}
\end{equation}
where $m_i~(i=1,2,\cdots, N)$ must be integers 
for $U(x^i, \tau)$ to satisfy the boundary condition (\ref{shiki3}).
\par
An important consequence of the gauge transformation (\ref{shiki4})
that depends on the Euclidean time coordinate linearly is to yield 
the shift symmetry for the zero mode of $A_{\tau}$,
\begin{equation}
A_{\tau}\rightarrow A_{\tau}^{\prime}=UA_{\tau}U^{\dagger}+{{2\pi T}\over g}
\left(\begin{array}{cccc}
m_1 & & & \\
    &m_2 & &  \\
 & & \ddots & \\
& & & m_N \\
\end{array}
\right).
\label{shiki5}
\end{equation}
%
%
Writing 
\begin{equation}
{g\over T}\vev{A_{\tau}}
={\rm diag}.\left(\varphi_1, \varphi_2,\cdots, \varphi_N\right)
\quad {\rm with}
\quad
\sum_{i=1}^N\varphi_i=0,
\label{shiki6}
\end{equation}
we find that
\begin{equation}
{g\over T} \vev{A_{\tau}^{\prime}}={\rm diag}.
\left(\varphi_1+2\pi m_1, \varphi_2+2\pi m_2,\cdots, \varphi_N+2\pi m_N\right).
\label{shiki7}
\end{equation}
We observe that if $\varphi_i$ is an integral multiple of $2\pi$, by
choosing appropriate values for
$m_i$, the configuration $\vev{A_{\tau}^{\prime}}$ is gauge equivalent to 
$\vev{A_{\tau}}=0$. On the other hand, if $\varphi_i$ is not an integral
multiple of $2\pi$, the configuration is 
physically distinct from $\vev{A_{\tau}}=0$. Hence, one
concludes that the VEV cannot be gauged away and is a dynamical variable.
\par
A physical consequence of the nontrivial VEV of $A_{\tau}$ is to
make the gauge bosons massive through the coupling
%
%
%
\begin{equation}
g^2{\rm tr}\left[\vev{A_{\tau}},~~A_i\right]^2.
\label{shiki8}
\end{equation}
The massive gauge boson is a signal for gauge symmetry 
breaking, so that the gauge symmetry breaking patterns are determined 
by $\vev{A_{\tau}}$.
\par
\section{Effective potential for $\vev{A_{\tau}}$}
Once we understand that $\vev{A_{\tau}}$ is a dynamical variable, we 
evaluate the effective potential for $\vev{A_{\tau}}$ in order 
to determine its value. Let us resort to perturbation theory in 
the one-loop approximation. 
Then, the VEV is determined as the minimum of the effective
potential. We expand the gauge field around the VEV as 
\begin{equation}
A_{\hmu}=\vev{A_{\tau}}\delta_{\tau\hmu}+{\bar A}_{\hmu},
\label{shiki9}
\end{equation}
where $\vev{A_{\tau}}$ is given in Eq.(\ref{shiki6}). In addition
to the well-known one-loop effective potential arising from the pure
Yang-Mills theory, we newly present explicit forms of the 
one-loop effective potential coming from the massless/massive
matter.
\par
Following the standard prescription to evaluate the effective 
potential \cite{gross}, we obtain that
\begin{equation}
V_{gauge}^{T}=-{\Gamma(D/2)\over \pi^{D/2}}(D-2)T^D
\sum_{i,j=1}^N\sum_{n=1}^{\infty}
{1\over n^D}\cos\left[n(\varphi_i - \varphi_j)\right],
\label{shiki10}
\end{equation}
where we have ignored the divergences independent of the order
parameters \footnote{The effective potential is finite in the
sense that any divergence which depends on the order parameters
does not appear \cite{masiero}. It is said that the shift 
symmetry (\ref{shiki5}) is crucial for the finiteness.}. 
This is the contribution from the gauge sector to the effective
potential. The factor $D-2$ in Eq.(\ref{shiki10}) stands for the
on-shell degrees of freedom for the $D$ dimensional gauge 
field. It is easy to minimize the effective potential (\ref{shiki10}).
By taking account of the minus sign in the right hand side 
of Eq. (\ref{shiki10}), the minimum of
the effective potential is given by
\begin{equation}
\varphi_i= {2\pi k\over N},~~(k=0,1,2,\cdots, N-1),\quad 
\varphi_N=-{2\pi k\over N}(N-1)={2\pi k\over N}~({\rm mod}~~2\pi).
\label{shiki11}
\end{equation}
Although there are $N$ distinct vacuum configurations labeled by
the integer $k$, they are physically equivalent. To see this, let 
us consider the Polyakov loop defined by
\begin{equation}
W_p\equiv {\cal P}~{\rm exp}\left({ig\int_0^{\beta}d\tau~\vev{A_{\tau}}}\right)
=\e^{i{2\pi k\over N}}{\bf 1}_{N\times N}.
\label{shiki12}
\end{equation}
The vacuum configuration takes the values at the center 
of $SU(N)$ for the VEV (\ref{shiki11}). In particular, for 
$N=3$  the vacuum is $Z_3$ symmetric, and the result is 
consistent with the lattice 
calculation \cite{iwasaki} in the high temperature region. The 
commutator in Eq. (\ref{shiki8}) vanishes for the configuration 
(\ref{shiki11}). The gauge boson remains massless, so that the gauge 
symmetry is not broken through the VEV of $A_{\tau}$.
\par
Let us next introduce fermions into the theory and study the vacuum
structure. We first consider the fundamental fermion under the gauge
group $SU(N)$. As stated in the introduction, the boundary condition 
of the fermion for the Euclidean time direction is fixed to be 
antiperiodic because of the quantum statistics, which is a big
difference from the case of spatial
extra dimensions, where one does not know, {\it a priori}, the boundary
condition for the spatial compactified direction. We must impose 
\begin{equation}
\psi(\tau+\beta)=-\psi(\tau).
\label{shiki13}
\end{equation}
The contribution from the fundamental fermion to the effective potential
in the one-loop approximation is given by
\begin{equation}
V_{fd}^{f,T}={\Gamma(D/2)\over \pi^{D/2}}2^{[D/2]}
T^D\sum_{i=1}^N\sum_{n=1}^{\infty}{1\over n^D} 
\cos\left[n(\varphi_i-\pi)\right],
\label{shiki14}
\end{equation}
where the factor $2^{[D/2]}$ stands for the on-shell degrees of freedom
of the fermion in $D$ dimensions. The shift of the argument 
in Eq.(\ref{shiki14}) by an amount of $\pi$ reflects the antiperiodic 
boundary condition (\ref{shiki13}).
\par
For $D=4$, by using the formula \footnote{The behavior of 
$\sum_{n=1}^{\infty}(-1)^n\cos(nx)/n^D (-\pi\leq x\leq \pi)$ is 
almost similar to $-\cos(x)$ for any $D\geq 3$. And the location of
the minimum of the function is $x=0$.},
\begin{equation}
\sum_{n=1}^{\infty}{(-1)^n \over n^4}\cos(nx)
=-{1\over 48}\left(x^2 -\pi^2\right)^2 +{\pi^4\over 90}\qquad 
(-\pi \leq x \leq \pi),
\label{shiki15}
\end{equation}
whose minimum is $x=0$, we find that the configuration
that minimizes $V_{fd}^{f,T}$ is given by 
\begin{equation} 
\varphi_i=0\qquad (i=1,2,\cdots, N).
\label{shiki16}
\end{equation}
\par
Let us next introduce the adjoint fermion under the gauge group $SU(N)$.
As before, the boundary condition of the fermion for the Euclidean 
time direction must be antiperiodic. The contribution to the effective 
potential is given by
\begin{equation}
V_{adj}^{f,T}={\Gamma(D/2)\over \pi^{D/2}}2^{[D/2]}
T^D\sum_{i,j=1}^N\sum_{n=1}^{\infty}{1\over n^D} 
\cos\left[n(\varphi_i-\varphi_j-\pi)\right].
\label{shiki17}
\end{equation}
In case of $D=4$, by using the formula (\ref{shiki15}), we find 
that the minimum of the effective potential (\ref{shiki17}) is given by
Eq.(\ref{shiki11}). This result also holds for $D\geq 5$. 
\par
Let us study the contributions from bosons instead of the 
fermions. Due to the Bose statistics, they obey the
periodic boundary condition. The contribution to the effective 
potential from the adjoint and fundamental bosons under 
the gauge group $SU(N)$ are given by
\begin{eqnarray}
V_{adj}^{b,T}&=&-{\Gamma(D/2)\over \pi^{D/2}}
T^D\sum_{i,j=1}^N\sum_{n=1}^{\infty}{1\over n^D} 
\cos\left[n(\varphi_i-\varphi_j)\right],
\label{shiki18}\\
V_{fd}^{b,T}&=&-2{\Gamma(D/2)\over \pi^{D/2}}
T^D\sum_{i=1}^N\sum_{n=1}^{\infty}{1\over n^D} 
\cos\left(n\varphi_i\right),
\label{shiki19}
\end{eqnarray}
respectively. $V_{adj}^{b,T}$ is the same form 
as $V_{gauge}^T$ aside from the on-shell
degrees of freedom. The configuration that minimizes 
$V_{adj}^{b,T}$ is given by
Eq.(\ref{shiki11}). By noting the minus sign 
in Eq.(\ref{shiki19}), the minimum of the potential
$V_{fd}^{b,T}$ is given by
\begin{equation}
\varphi_i= 0\quad (i=1,2,\cdots, N).
\label{shiki20}
\end{equation}
\par
We are now in a position to discuss a general property of the
total effective potential,
\begin{equation}
V_{total}^T=V_{gauge}^T+N_{fd}^fV_{fd}^{f,
T}+N_{adj}^fV_{adj}^{f, T}
+N_{fd}^bV_{fd}^{b,T}+N_{adj}^bV_{adj}^{b,T},
\label{shiki21}
\end{equation}
where $N_{fd}^f$ and $N_{adj}^f$ $(N_{fd}^b~{\rm and}~N_{adj}^b)$
are the numbers of the fundamental and the adjoint fermions
(bosons), respectively. We have shown that each of the potentials 
$V_{gauge}^T, V_{adj}^{f, T}$ and $V_{adj}^{b, T}$ has a
minimum at $\varphi_i=2\pi k/N~{\rm mod}~2\pi~(i=1,2,\cdots, N)$  
for $k=0,1,\cdots, N-1$, and that each of the potentials
$V_{fd}^{f, T}$ and $V_{fd}^{b, T}$ has a minimum 
at $\varphi_i=0~(i=1,2,\cdots, N)$. These results immediately
imply that the vacuum configuration of $\varphi_i$ is given by
\begin{equation}
\varphi_i=\left\{
\begin{array}{ll}
{2\pi k\over N} \ \ ({\rm mod}\ 2\pi)
&\ {\rm for}~~N_{fd}^f=N_{fd}^b=0,\\[0.3cm]
\ 0 &\ {\rm for~~otherwise},\ \ \ \ ~(i=1,2,\cdots,N).
\end{array}
\right.
\label{shiki22}
\end{equation}
Thus, we conclude that nontrivial values of $\vev{A_{\tau}}$
are not realized at finite temperature and the gauge symmetry
cannot be broken for the $SU(N)$ gauge theory with arbitrary
numbers of the fermions and the bosons belonging to the
fundamental and the adjoint representations.
\par
In all cases we have considered above, the gauge symmetry is not broken 
by the values of $\varphi_i$. The result does not
depend on the space-time dimensions $D (\geq 3)$. It should be noted that the
boundary condition for the Euclidean time direction, which is uniquely 
fixed by the quantum statistics, is crucial for the results we
have obtained. To see this, let us relax the boundary
condition for the Euclidean time direction. Then, the contribution from the
adjoint fermion to the effective potential is 
\begin{equation}     
V_{adj}^{relaxed,f,T}={\Gamma(D/2)\over \pi^{D/2}}2^{[D/2]}
T^D\sum_{i,j=1}^N\sum_{n=1}^{\infty}{1\over n^D} 
\cos\left[n(\varphi_i-\varphi_j-\alpha)\right],
\label{shiki23}
\end{equation}
where $\alpha$ parametrizes the boundary condition for the Euclidean
time direction
\begin{equation}
\psi(\tau +\beta)=\e^{i\alpha}~\psi(\tau).
\label{shiki24}
\end{equation}
The configuration that minimizes
the potential $V_{adj}^{relaxed,T}$ depends on $\alpha$, and it is given  
for $\alpha=0, D=4$ by 
\begin{equation}
{g\over T}\vev{A_{\tau}}=
\left({N-1\over N}, {N-3\over N}, \cdots,0,\cdots, {-N+3\over N},
{-N+1\over N}\right).
\label{shiki25}
\end{equation} 
This configuration breaks the $SU(N)$ symmetry down to its maximally
abelian subgroup $U(1)^{N-1}$ \cite{abelian}.
\par
We have, so far, considered the adjoint and fundamental matter. In
order to see how the VEV is affected by the
representation under the gauge group, let us study higher
dimensional representations of $SU(2)$ as an example. We evaluate the 
contribution to the effective potential from the fermion belonging 
to the spin-$j$ representation. 
\par
We take the Cartan matrix $T^3$ for the 
spin-$j$ representation of the $SU(2)$ as
\begin{equation}
\left(T^3 \right)_{(2j+1)\times (2j+1)}
={\rm diag}.\left(j, j-1, j-2,\cdots, -(j-2), -(j-1),-j\right).
\label{shiki26}
\end{equation}
The covariant derivative in this case is given by
\begin{equation}
(D_{\tau})_{ab}=\delta_{ab}\del_{\tau} -ig \vev{A_{\tau}^3}(T^3)_{ab}
=\delta_{ab}\del_{\tau} -ig {2T\varphi\over g}(T^3)_{ab},
\label{shiki27}
\end{equation}
where we have used the equation (\ref{shiki6}) with $N=2$ and the
indices $a, b$ run from $1$ to $2j+1$. Then, the 
eigenvalues for the mass operator $(D_{\tau})^2$ in momentum space is
\begin{eqnarray}
-(D_{\tau})^2&=&(2\pi T)^2{\rm diag}.\nonumber\\
&\times& \left(
(n-{2\varphi j\over 2\pi})^2, 
(n-{2\varphi (j-1)\over 2\pi})^2, \cdots, 
(n+{2\varphi (j-1)\over 2\pi})^2, 
(n+{2\varphi j\over 2\pi})^2 \right).
\label{shiki28}
\end{eqnarray}
Hence, the contribution to the effective potential is given by
\begin{equation}
V_{spin-j}^{T}={\Gamma(D/2)\over \pi^{D/2}}2^{[D/2]}T^D
\sum_{n=1}^{\infty}\sum_Q
{2\over n^D}\cos\left[n(2Q\varphi -\pi)\right],
\label{shiki29}
\end{equation}
where $Q=1,2,\cdots, j$ for $j\in {\bf Z}, Q=1/2,3/2,\cdots,j$
for $j\in {\bf Z}+1/2$, and we have 
taken account of the Fermi statistics for the boundary
condition for the Euclidean time direction. If we set $D=4$ and 
apply the formula (\ref{shiki15}), we find that
the minimum of Eq. (\ref{shiki29}) is realized 
by $\varphi=0~(\varphi=0,~\pi)$ 
for $j\in {\bf Z}+1/2~(j\in {\bf Z})$. Hence, the vacuum 
configuration for the total effective 
potential $V_{gauge}^{T}+V_{spin-j}^T$ is given by
$\varphi=0~(\varphi=0,~\pi)$ for $j={\bf Z}+1/2~(j\in{\bf Z})$,
so that the $SU(2)$ gauge symmetry is not broken.
\par
One may wonder the vacuum configuration changes if we consider
massive instead of the massless matter. In case of the massive
matter, the contribution to the effective potential 
becomes 
\begin{equation}
V_{massive}^T=N_{deg}(-1)^{f+1}{2\over(2\pi)^{D\over 2}}
\sum_{i=1}^N\sum_{n=1}^{\infty}
\left(M^2\over(n/T)^2\right)^{D\over 4}
K_{D \over 2}\left(M{n\over T}\right)\cos[n(\varphi_i-2\pi\eta)]
\label{shiki30}
\end{equation}
for matter belonging to the fundamental representation under 
the $SU(N)$. Here $M$ denotes the bulk mass for the matter. $f$ stands for the 
fermion number and $\eta=\half (0)$ for fermions (bosons). $N_{deg}$ denotes 
the on-shell degrees of freedom and $K_{D\over 2}$ is the modified 
Bessel function. For the adjoint matter, it is understood that 
$\varphi_i\rightarrow \varphi_i-\varphi_j, \sum_i\rightarrow
\sum_{i, j}$ in Eq.(\ref{shiki30}). It 
is known that for $D/2=$ half integer, the modified Bessel function 
is written in terms of the elementary function, for example, 
\begin{equation}
K_{5\over 2}(z)=3\sqrt{{\pi\over 2 z^5}}\left(1+z+{z^2\over 3}\right)\e^{-z}.
\label{shiki31}
\end{equation}
One recovers the former 
results (\ref{shiki14}) and (\ref{shiki19}) by noting 
that\footnote{The equation (\ref{shiki32}) holds for any $D$.}  
\begin{equation}
\lim_{M\rightarrow 0}K_{D\over 2}(aM)={1\over (aM)^{D\over 2}}
2^{{D\over 2}-1}\Gamma(D/2),
\label{shiki32}
\end{equation}
where $a\equiv n/T$. As seen in Eq.(\ref{shiki31}), heavy 
particles tend to decouple from the effective
potential. This is nothing but the Boltzmann suppression which states
that particles with smaller wavelengths than the inverse temperature 
has exponentially suppressed distribution in the system.
\par
For $D=5$, the equation (\ref{shiki30}) becomes
\begin{eqnarray}
V_{massive}^T&=&N_{deg}(-1)^{f+1}\left({3\over{4\pi^2}}\right)
T^5\sum_{i=1}^N\sum_{n=1}^{\infty}{1\over n^5}
\left(
1+n{M\over T}+{1\over 3}{n^2M^2\over T^2}
\right)\e^{-n{M\over T}}\nonumber\\
&&\times \cos[n(\varphi_i-2\pi\eta)].
\label{shiki33}
\end{eqnarray}
When $\omega\equiv M/T\ll 1, \varphi_i\ll 1$, the expansion 
formula is known \cite{formula},
\begin{eqnarray}
&&(-1)^{f+1}\sum_{n=1}^{\infty}{1\over n^5}
\left(1+n\omega+{n^2\omega^2\over 3}\right)
\e^{-n\omega}\cos[n(\varphi-2\pi\eta)] \nonumber\\
&=&\left\{
\begin{array}{lll}
-\zeta(5)+{\zeta(3)\over 6}\omega^2-{\omega^4\over 32}+
{\zeta(3)\over 2}\varphi^2 -{7\over 48}\omega^2\varphi^2
+O(\varphi^4)& \mbox{for} & \mbox{boson},
\\[0.3cm]
-{15\over 16}\zeta(5)+{\zeta(3)\over 8}\omega^2
+{3\zeta(3)\over 8}\varphi^2 
-{{\rm ln}2\over 12}\omega^2\varphi^2
+O(\varphi^4) 
& \mbox{for} & \mbox{fermion}.
\end{array}\right.
\label{shiki34}
\end{eqnarray}
We observe that the bulk mass tends to make the curvature of the
effective potential at the origin negative, so that the 
nontrivial value for $\varphi$ would be induced due to 
the presence of the massive bulk fermion. 
\par
If we consider the 
fundamental fermion under $SU(N)$, for instance, the coefficient
of the quadratic term of the total effective potential 
is given, aside from the overall factor $T^{-5}(3/4\pi^2)^{-1}$, by 
\begin{equation}
12\zeta(3)+8N_{fd}\left({3\over 8}\zeta(3)-{{\rm ln}2\over 12}\omega^2\right),
\label{shiki35}
\end{equation}
where the first term comes from the gauge sector in the effective
potential and $N_{fd}$ is the number of the fundamental fermions.
The mass term, however, cannot be negative for the small
bulk mass $\omega \ll 1$ because of the large contribution, the first
and the second terms in Eq.(\ref{shiki35}). 
Therefore, the bulk mass does not lead to nontrivial values of
$\varphi$, and hence, the gauge symmetry is not broken.
\par
In this section, we have calculated the one-loop effective
potentials of $\vev{A_{\tau}}$ for the $SU(N)$ gauge theory
with/without the (massless or massive) matter belonging to the
various representations. Our results show that nontrivial values
for $\vev{A_{\tau}}$ are not realized and the gauge symmetry
cannot be broken. 
\section{Gauge theories compactified on $S^1$ at finite
temperature}
We have seen that no nontrivial values for $A_{\tau}$ are
obtained for the finite temperature case and have understood the
boundary condition for the Euclidean time direction is crucial. In 
this section we investigate further the possibility that 
the $\vev{A_{\tau}}$ takes nontrivial value at finite temperature. 
To this end, we consider one spatial extra dimension which is 
compactified on $S^1$ whose radius is $R$.
The coordinate of the
$S^1$ is denoted by $y$, so that our space-time 
in this case is $S^1_{\tau}\times R^{D-2}\times S^1$. The gauge potential is 
decomposed as $(A_{\tau},A_k, A_y)~(k=1,\cdots, D-2)$. $A_{\tau}$ and
$A_y$ are dynamical degrees of freedom corresponding to the two
independent circles.
\par
Let us note that the tree-level potential from 
the coupling ${\rm tr}F_{y\tau}^2$ arises as 
\begin{equation}
V_{tree}= g^2\,{\rm tr}\left[\vev{A_{\tau}},~\vev{A_{y}}\right]^2.
\label{shiki36}
\end{equation}
Writing 
\begin{equation}
\vev{A_y}={1\over gL}\vev{{\tilde A}_y},\quad 
\vev{A_{\tau}}={T\over g}\vev{{\tilde A}_{\tau}},
\label{shiki37}
\end{equation}
we have 
\begin{equation}
V_{tree}={T^2\over L^2 g^2}\,{\rm tr}
\left[\vev{{\tilde A}_{\tau}},~\vev{{\tilde A}_{y}}\right]^2,
\label{shiki38}
\end{equation}
where $L\equiv 2\pi R$. 
%
%
We observe that in the weak 
coupling limit the tree-level potential dominates, so that it is 
natural to expect that the
vacuum configuration lies along the flat direction,
\begin{equation}
\left[\vev{{\tilde A}_{\tau}},~\vev{{\tilde A}_{y}}\right]=0.
\label{shiki39}
\end{equation}
By utilizing the gauge degrees of freedom, we can 
take, for example, $\vev{{\tilde A}_y}$ in diagonal form, so that from
Eq.(\ref{shiki39}), we parametrize the VEVs as 
\begin{equation}
\vev{{\tilde A}_{\tau}}={\rm diag}.
\left(\varphi_1, \cdots, \varphi_N\right),
\qquad
\vev{{\tilde A}_{y}}={\rm diag}.
\left(\theta_1, \cdots, \theta_N\right),
\label{shiki40}
\end{equation}
where $\sum_{i=1}^{N}\theta_i = \sum_{i=1}^{N}\varphi_i=0$. 
In the background (\ref{shiki40}),  we have the 
quadratic terms with respect to the
fluctuation, $g^2([\vev{A_{\tau}},~A_k]^2+[\vev{A_y},~A_k]^2)$, which, 
by expanding the $A_k$ in terms of the Fourier modes, yield
the gauge boson mass for $(A_k^{({\bar n},n)})_{ij}$,
\begin{equation}
(2\pi T)^2\left({\bar n}+{{\varphi_i-\varphi_j}\over 2\pi}\right)^2
+{1\over R^2}\left(n+{{\theta_i-\theta_j}\over 2\pi}\right)^2,
\label{shiki41}
\end{equation}
where the integer $n ({\bar n})$ is the Kaluza-Klein (Matsubara) mode.
We see that the nontrivial VEVs are the signal for gauge
symmetry breaking. 
\par
We calculate the effective potential for $\varphi_i$'s and $\theta_i$'s
according to the standard prescription. We find 
that $V_{eff}=N_{deg}F$, where
\begin{eqnarray}
F&=&(-1)^{f+1}
{2\over(2\pi)^{D\over 2}}
\Biggl[
\sum_{i=1}^{N}\sum_{m=1}^{\infty}
\left(M^2\over (Lm)^2\right)^{D\over 4}
K_{D\over 2}(LMm)\cos\left[m (\theta_i-\alpha)\right]
\nonumber\\
&&+\sum_{i=1}^{N}\sum_{l=1}^{\infty}
\left(M^2\over (l/T)^2\right)^{D\over 4}
K_{D\over 2}\left(M{l \over T}\right)\cos\left[l(\varphi_i+2\pi\eta)\right]
\nonumber\\
&&+2\sum_{i=1}^N\sum_{m,l=1}^{\infty}
\left({M^2\over{(Lm)^2+(l/T)^2}}\right)^{D\over 4}
K_{D\over 2}\left(\sqrt{(MLm)^2+(Ml/T)^2}\right)  \nonumber\\
&&\times 
\cos\left[m (\theta_i-\alpha)\right]
\cos\left[l(\varphi_i+2\pi\eta) \right]
\Biggr].
\label{shiki42}
\end{eqnarray}
Here we have considered the matter belonging to the fundamental
representation with the bulk mass. $N_{deg}=2^{[D/2]}N_{fd}$
stands for the on-shell degrees of freedom with the flavor number
$N_{fd}$, and $M$ denotes 
the bulk mass for the matter. It should be emphasized that one
does not know, {\it a priori}, the boundary condition for
the spatial compactified direction. The parameter $\alpha$ comes 
from the twisted boundary condition 
for the $S^1$ direction (the spatial extra dimension),
\begin{equation}
\phi_{matter}(y+L)=\e^{i\alpha}\phi_{matter}(y).
\label{shiki43}
\end{equation}
$\eta$ takes $1/2 (0)$ for the fermion (boson). In case 
of the adjoint matter, the argument of the cosine 
function and the summation should be replaced by 
\begin{eqnarray}
&&\varphi_i \rightarrow \varphi_i-\varphi_j,
\qquad
\theta_i \rightarrow \theta_i-\theta_j,\nonumber\\
&&
\sum_i\rightarrow \sum_{i, j}.
\label{shiki44}
\end{eqnarray}
The contribution from the gauge sector is reproduced by $\alpha=\eta=M=0$
in Eq.(\ref{shiki42}).
\par
Let us also note that the potential (\ref{shiki42}) 
with $\alpha=0, f=0~(\eta=0)$ or $\alpha=\pi, f=1~(\eta=1/2)$ 
is invariant under the exchanges 
\begin{equation}
L\leftrightarrow T^{-1}
\ \  {\rm and }\ \ \theta_i \leftrightarrow \varphi_i\, .  
\label{shiki45}
\end{equation}
If we define the dimensionless 
parameters $t\equiv LT$ and $z\equiv ML$,
the equation (\ref{shiki42}) becomes
\begin{eqnarray}
F&=&(-1)^{f+1}
{2\over(2\pi)^{D\over 2}}{1\over L^D}
\Biggl[
\sum_{i=1}^{N}\sum_{m=1}^{\infty}
\left(z^2\over m^2\right)^{D\over 4}
K_{D\over 2}(mz)\cos\left[m (\theta_i-\alpha)\right]
\nonumber\\
&&+t^D\sum_{i=1}^{N}\sum_{l=1}^{\infty}
\left((z/t)^2\over l^2\right)^{D\over 4}
K_{D\over 2}\left({zl \over t}\right)
\cos\left[l(\varphi_i+2\pi\eta)\right]
\nonumber\\
&&+2t^D\sum_{i=1}^N\sum_{m,l=1}^{\infty}
\left({(z/t)^2\over{(mt)^2+l^2}}\right)^{D\over 4}
K_{D\over 2}\left(\sqrt{(mz)^2+(zl/t)^2}\right)  \nonumber\\
&&\times 
\cos\left[m (\theta_i-\alpha)\right]
\cos\left[l(\varphi_i+2\pi\eta) \right]
\Biggr].
\label{shiki46}
\end{eqnarray}
\par
Let us comment on the effective 
potential (\ref{shiki42}) (or (\ref{shiki46})). We can obtain the
effective potential (\ref{shiki42}) if we consider two spatial
extra dimensions $T^2=S^1\times S^1$ at zero temperature 
and the boundary conditions
for one of the two $S^1$'s are specified by the quantum
statistics. 
Hence, the potential is the one obtained for a special 
case in the possible boundary conditions for the two $S^1$ directions.
\par
It may be instructive here to study the dependence of the potential on
the dimensionless parameter $t$. For $t\ll 1$, that 
is, $T\ll L^{-1}$ the dominant contribution to the potential comes 
from the $T=0$ part of the potential, that is, the first term 
in Eq.(\ref{shiki46}). The $S^1_{\tau}$ direction is effectively 
uncompactified in this limit, which implies that $\varphi$ is no longer the
dynamical variable and the potential becomes insensitive to the values of
$\varphi$. Then, the effective potential tends to be flat along 
the $S^1_{\tau}$ direction. The gauge symmetry can be broken through 
nontrivial values of $\theta_i$ in the limit by the Hosotani 
mechanism. On the other hand, if $t\gg 1$, that is, $T \gg L^{-1}$, the 
second term in Eq.(\ref{shiki46}) becomes dominant
and the effective potential reduces to the one discussed in the
previous section. 
Hence, the gauge symmetry is not broken
because $\varphi_i$ turns out to have no nontrivial VEV,
as we
have studied in the previous section. In this limit the potential 
becomes flat along the $S^1$ direction. 
\subsection{Massless matter}
Let us first consider $M\rightarrow 0$ limit. Using the 
equation (\ref{shiki32}), we obtain from Eq.(\ref{shiki42}) that
\begin{eqnarray}
{\bar F}&\equiv &{F\over {\Gamma({D\over 2})/\pi^{D\over 2}L^D}}
\nonumber\\
&=&(-1)^{f+1}
\Biggl[
\sum_{i=1}^N\sum_{m=1}^{\infty}{1\over m^D}
\cos\left[m(\theta_i-\alpha)\right]
+(LT)^D\sum_{i=1}^{N}\sum_{l=1}^{\infty}{1\over l^D}
\cos\left[l(\varphi_i+2\pi\eta)\right]
\nonumber\\
&&+2(LT)^D
\sum_{i=1}^{N}\sum_{l,m=1}^{\infty}
{1\over{\left[(mLT)^2+l^2\right]^{D/2}}}
\cos\left[m(\theta_i-\alpha)\right]
\cos\left[l(\varphi_i+2\pi \eta)\right]
\Biggr].
\label{shiki47}
\end{eqnarray}
\par
The contribution ${\bar F}_{gauge}$ from the gauge 
sector is given by setting $N_{deg}=D-2, f=\alpha=\eta=0$ with the 
replacement (\ref{shiki44}). 
\begin{eqnarray}
&&{\bar F}_{gauge}\nonumber\\
&&=
-\Biggl[
\sum_{i,j=1}^N\sum_{m=1}^{\infty}{1\over m^D}
\cos\left[m(\theta_i-\theta_j)\right]
+(LT)^D\sum_{i,j=1}^{N}\sum_{l=1}^{\infty}{1\over l^D}
\cos\left[l(\varphi_i-\varphi_j)\right]
\nonumber\\
&&\ \ +2(LT)^D
\sum_{i,j}^{N}\sum_{l,m=1}^{\infty}
{{\cos\left[m(\theta_i-\theta_j)\right]
\cos\left[l(\varphi_{i}-\varphi_{j})\right]}
\over{\left[(mLT)^2+l^2\right]^{D/2}}}\Biggr].
\label{neweq1}
\end{eqnarray}
The overall minus sign in the potential
implies that the vacuum configuration is given by 
\begin{equation}
\varphi_i={2\pi k\over N}~~(k=0,1,\cdots,N-1)~({\rm mod}~2\pi),\quad 
\theta_i={2\pi k'\over N}~~(k'=0,1,\cdots,N-1)~({\rm mod}~2\pi).
\label{shiki48}
\end{equation}
The Polyakov loop defined by Eq.(\ref{shiki12}) and the Wilson loop
\begin{equation}
W\equiv {\cal P}{\rm exp}\left(ig\int_{S^1}dy~\vev{A_y}\right)
\label{shiki49}
\end{equation}
commute with each other for the configurations (\ref{shiki48}), and 
they take the values at the center of $SU(N)$. The gauge boson 
$A_k^{(0,0)}$ remains 
massless from Eq.(\ref{shiki41}) for the vacuum 
configuration (\ref{shiki48}). The gauge symmetry is not broken in this case.
\par
Let us next consider the fundamental fermion, for which the potential is
given by $N_{deg}=2^{[D/2]}N_{fd}, f=1, \eta=\half$. For 
concreteness, if we consider 
$D=5$ case and the gauge group $SU(2)$, then, the potential becomes
\begin{eqnarray}
{\bar F}_{fd}&=&2\sum_{m=1}^{\infty}{1\over m^5}
\cos(m\alpha)\cos(m\theta)
+2(LT)^5\Biggl(
\sum_{l=1}^{\infty}{(-1)^l\over l^5}\cos(l\varphi)
\nonumber\\
&&+\sum_{m,l=1}^{\infty}
{2(-1)^l\over{[(mLT)^2+l^2]^{5/2}}}\cos(m\alpha)\cos(m\theta)
\cos(l\varphi)\Biggr).
\label{shiki50}
\end{eqnarray}
We depict the behavior of the potential (\ref{shiki50}) for $\alpha=0,
LT=1.0$ in Fig.$1$ and observe that the minimum is given by
\begin{equation}
(\varphi,~\theta)=(0,~\pi)~~({\rm mod}~2\pi).
\label{shiki51}
\end{equation}
\begin{figure}[ht]
\begin{center}
\includegraphics[width=9cm,height=8cm,keepaspectratio]
{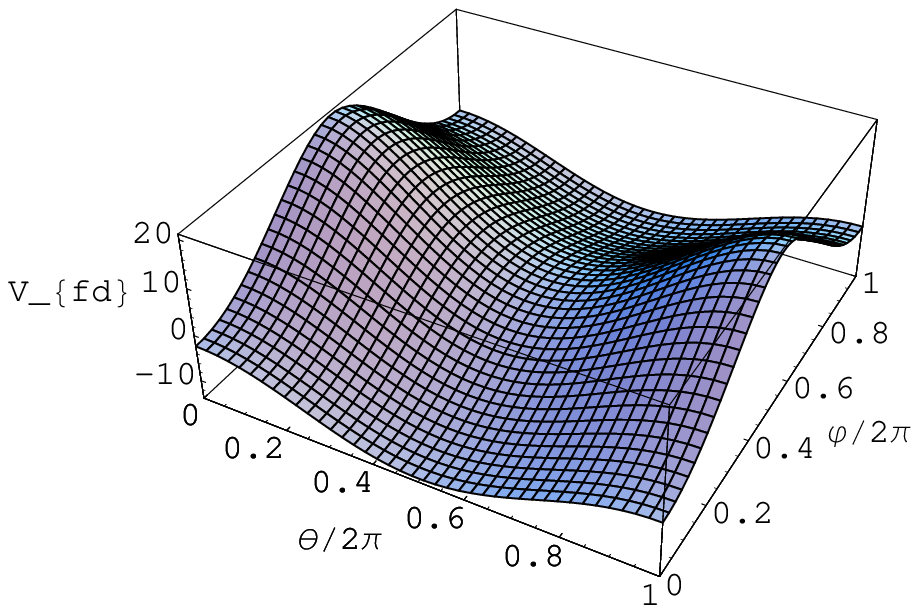}
\end{center}
\caption{The behavior of ${\bar F}_{fd}$ for $\alpha=0, LT=1.0$. The
minimum of the potential is given by $(\varphi,~\theta)
=(0,~\pi)~({\rm mod}~2\pi)$.}
\label{fig1}
\end{figure}
\par
For the adjoint fermion under the gauge group $SU(2)$, we obtain that
\begin{eqnarray}
{\bar F}_{adj}&=&2\sum_{m=1}^{\infty}{1\over m^5}
\cos(m\alpha)\bigl(1+\cos(2m\theta)\bigr)+
2(LT)^5
\Biggl(\sum_{l=1}^{\infty}{(-1)^l\over l^5}\bigl(1+\cos(2l\varphi)\bigr)
\nonumber\\
&&+\sum_{l,m=1}^{\infty}{{2(-1)^l}\over{[(mLT)^2+l^2]^{5/2}}}
\cos(m\alpha)\biggl(1+\cos(2m\theta)\cos(2l\varphi)\biggr)
\Biggr).
\label{shiki52}
\end{eqnarray}
From Fig.$2$, we observe that the minimum of the potential
(\ref{shiki52}) is given  by
\begin{equation}
(\varphi,~\theta)=\left(0,~{\pi\over 2}\right)~~({\rm mod}~\pi)
\label{shiki53}
\end{equation}
for $\alpha=0, LT=1.0$. 
\begin{figure}[ht]
\begin{center}
\includegraphics[width=9cm,height=8cm,keepaspectratio]
{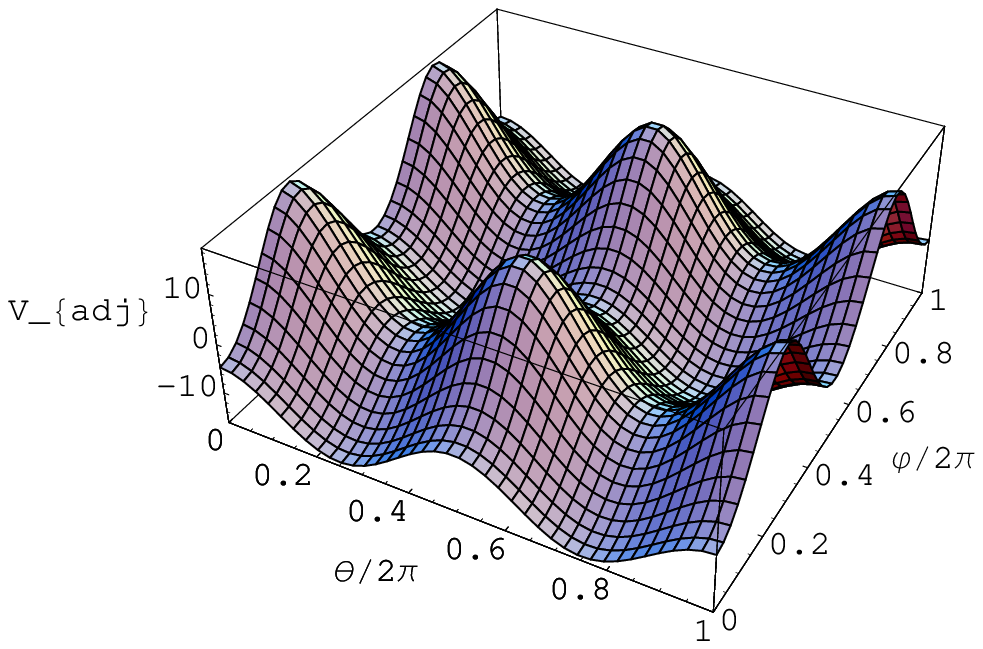}
\end{center}
\caption{The behavior of ${\bar F}_{adj}$ for $\alpha=0, LT=1.0$. The
minimum of the potential is given by $(\varphi,~\theta)
=(0,~\pi/2)~({\rm mod}~\pi)$.}
\label{fig2}
\end{figure}
\par
Let us now study the total system, that is, the fermions coupled 
to the gauge field. We first consider the fundamental 
fermion and the gauge field. The behavior of
the total effective potential 
$N_{gauge} {\bar F}_{gauge} + 4N_{fd} {\bar F}_{fd}$ is 
depicted in Fig.$3$, where we take $\alpha=0, LT=1.0$.
\begin{figure}[ht]
\begin{center}
\includegraphics[width=9cm,height=8cm,keepaspectratio]
{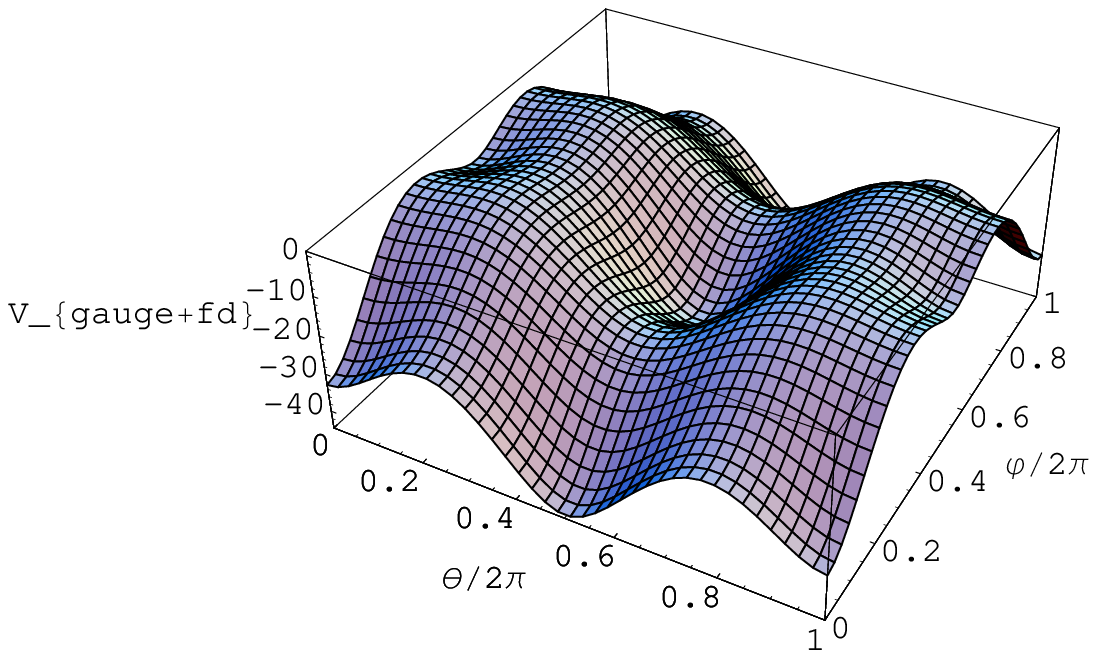}
\end{center}
\caption{The behavior of $N_{gauge} {\bar F}_{gauge}
+4N_{fd} {\bar F}_{fd}$ for $(N_{gauge},N_{fd})
=(3,1),\alpha=0, LT=1.0$. The minimum of the 
potential is given by $(\varphi,~\theta)
=(0,~\pi)~({\rm mod}~2\pi)$.}
\label{fig3}
\end{figure}
The minimum of the total effective potential is given by
\begin{equation}
(\varphi,~\theta)=(0,~\pi)~~({\rm mod}~2\pi).
\label{shiki55}
\end{equation}
$A_k^{(0,-1)}$ is still a massless mode for the vacuum 
configuration from Eq.(\ref{shiki41}), so that 
the $SU(2)$ gauge symmetry is not broken in this case.
\par
The fermion contribution to the effective potential depends on the
parameter $\alpha$, which twists the boundary condition for the
$S^1$ direction (the spatial extra dimension). The physical region 
of $\alpha$ is given by $0\leq \alpha\leq \pi$. Since the effect 
of $\alpha$ shifts only the $\theta$, as seen in Eq.(\ref{shiki42}), the 
configuration that minimizes (\ref{shiki50}) changes according 
to $\alpha$. We numerically confirm that the vacuum configuration 
for the total potential 
$N_{gauge} {\bar F}_{gauge} + 4N_{fd} {\bar F}_{fd}$ 
is given by
\begin{equation}
(\varphi,~\theta)=\left\{
\begin{array}{lll}
(0,~\pi) & \mbox{for}  & 0\leq \alpha < {\pi\over 2},\\[0.3cm]
%
%
(0,~0) & \mbox{for} & {\pi\over 2}\leq  \alpha \leq \pi.
\end{array}\right.
\label{shiki56}
\end{equation}
Let us note that at $\alpha_c\equiv \pi/2$ the fermion 
contribution (\ref{shiki50}) is invariant under the 
translation $\theta\rightarrow \theta+\pi$; that
is, the periodicity with respect to $\theta$ becomes half of the
original periodicity $2\pi$, so that if $\theta=0$ is the minimum 
configuration, so is $\theta=\pi$. The 
result (\ref{shiki56}) and the critical value $\alpha_c$ are 
independent of the flavor number of the fundamental fermions. 
The order parameter $\varphi$ does not take nontrivial values in this case. 
Note that, even though $\theta=\pi$, the mode $A_k^{(0,-1)}$ is a
massless mode, so that the $SU(2)$ gauge symmetry is not broken.  
\par
Let us next consider the adjoint fermion instead of the fundamental
fermion, whose contribution to the effective potential is given by 
Eq.({\ref{shiki52}}). The behavior of the total potential
$N_{gauge} {\bar F}_{gauge} + 4N_{adj} {\bar F}_{adj}$ 
is shown in Fig.$4$ for $LT=1.0, \alpha=0$. 
\begin{figure}[ht]
\begin{center}
\includegraphics[width=9cm,height=8cm,keepaspectratio]
{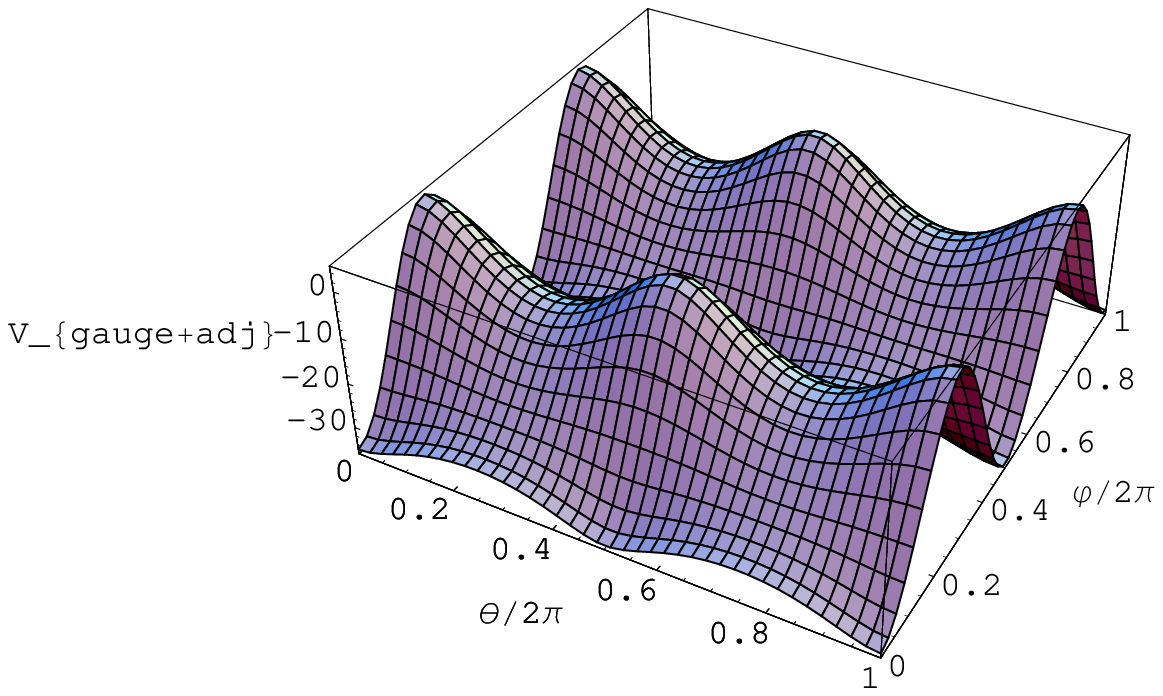}
\end{center}
\caption{The behavior of $N_{gauge}{\bar F}_{gauge}+
4N_{adj}{\bar F}_{adj}$ 
for $(N_{gauge},N_{adj})=(3,1),\alpha=0, LT=1.0$. The 
minimum of the potential is 
given by $(\varphi,~\theta)=(0,~0)~({\rm mod}~\pi)$.}
\label{fig4}
\end{figure}
The vacuum configuration is 
\begin{equation}
(\varphi,~\theta)=(0,~0)~(\mbox{mod}~\pi).
\label{shiki57}
\end{equation}
The $SU(2)$ gauge symmetry is not broken in this case.
\par
Let us consider the effect of $\alpha$
and the number of the adjoint fermions on the vacuum configuration. We 
numerically confirm that 
for $N_{adj}=1$, where $N_{adj}$ is the number of the adjoint
fermions, the vacuum 
configuration $(\varphi,~\theta)=(0,~0)$ is independent of the values of
$\alpha$. If $N_{adj}\geq 2$, we find critical values 
of $\alpha$, above and  below which the vacuum configuration is different,
\begin{equation}
(\varphi,~\theta)=
\left\{
\begin{array}{lll}
(0,~\pi/2) & \mbox{for}& 0 \leq \alpha < {\alpha_c},\\[0.3cm]
(0,~0) & \mbox{for} & \alpha_c < \alpha < \pi,
\end{array}\right.
\label{shiki58}
\end{equation}
where, for example, $\alpha_c/2\pi
\sim 0.1063, 0.1623, 0.1857, 0.1991, 0.2077$ for 
$N_{adj}=2, 3, 4, 5, 6$, respectively. 
$\varphi$ does not take nontrivial 
values in this case. The gauge 
boson becomes massive only through the nontrivial values 
of $\theta$, and the $SU(2)$ gauge symmetry is broken 
down to $U(1)$ by the Hosotani mechanism.
\par
Let us finally consider both the adjoint and fundamental fermions.
The effective potential is given by 
$N_{gauge}{\bar F}_{gauge}+4N_{fd}{\bar F}_{fd}+
4N_{adj}{\bar F}_{adj}$. From the lessons
obtained above, we expect that $\varphi$ does not take any 
nontrivial values, while $\theta$ depends on the parameter $\alpha$ and 
the number of flavor introduced in the theory. As an illustration, let us 
choose $(N_{fd},~N_{adj})=(1,~5)$ and $\alpha=0$. The behavior of the
effective potential is given in Fig.$5$, where we observe that the
vacuum configuration is $(\varphi,~\theta)=(0,~0.261)\times 2\pi$.
\begin{figure}[ht]
\begin{center}
\includegraphics[width=9cm,height=8cm,keepaspectratio]
{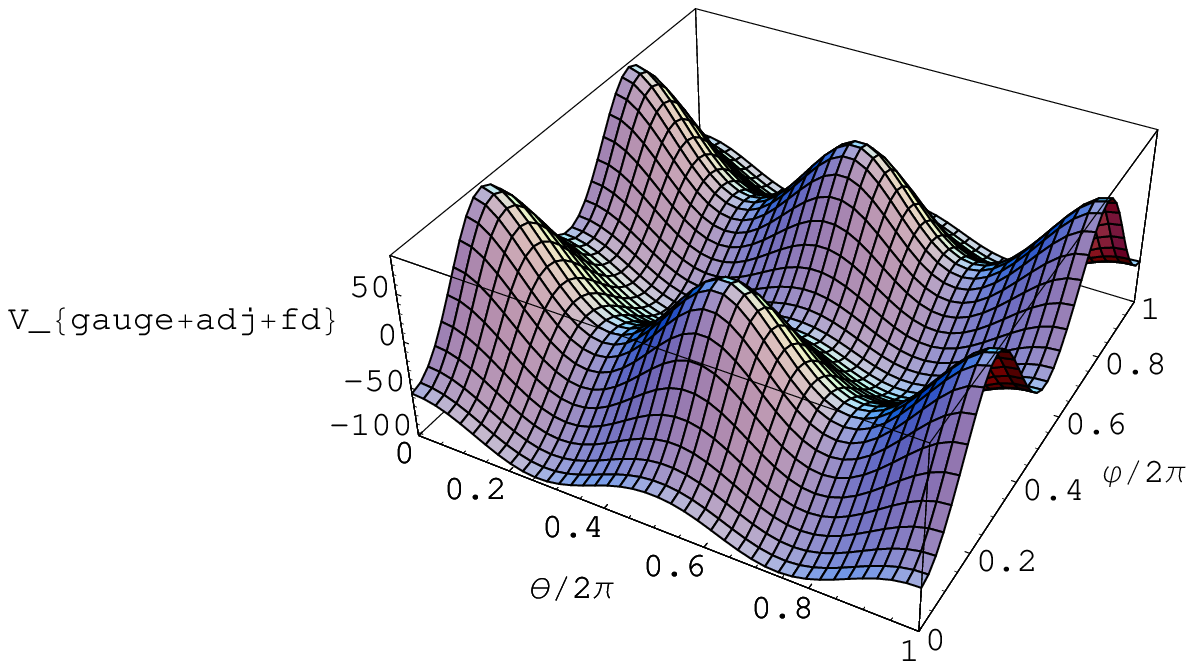}
\end{center}
\caption{The behavior of $N_{gauge}{\bar F}_{gauge}+
4N_{fd}{\bar F}_{fd}+4N_{adj}{\bar F}_{adj}$ for 
$(N_{gauge},N_{fd},N_{adj})=(3,1,5), \alpha=0, LT=1.0$. The minimum 
of the potential is given by $(\varphi,~\theta)
=(0,~0.262)\times 2\pi~({\rm mod}~2\pi)$.}
\label{fig5}
\end{figure}
If we take $\alpha/2\pi=0.2$ for the same matter content, the vacuum 
configuration changes to $(\varphi,~\theta)=(0.0,~0.5)\times 2\pi$. 
\par
\subsection{Massive matter}
In this subsection we study the effect of the massive bulk fermions on 
the vacuum configuration. We consider the five dimensional case. By using 
the formula (\ref{shiki31}), the equation (\ref{shiki46}) becomes
\begin{eqnarray}
L^5\left({3\over4\pi^2}\right)^{-1}F
&=&
(-1)^{f+1}
\Biggl[
\sum_{i=1}^N\sum_{m=1}^{\infty}{1\over m^5}
\left(1+mz+{m^2z^2\over 3}\right)\e^{-mz}\cos[m(\theta_i-\alpha)]
\nonumber\\
&&+t^5\sum_{i=1}^N\sum_{l=1}^{\infty}{(-1)^l\over l^5}
\left(1+{lz\over t}+{l^2z^2\over 3~t^2}\right)\e^{-{lz/t}}
\cos(l(\varphi_i+2\pi\eta))\nonumber\\
&&+2t^5\sum_{i=1}^N\sum_{l,m=1}^{\infty}
{(-1)^l\over {[(mt)^2+l^2]^{5/2}}}\nonumber\\
&&\times \left(1+\sqrt{(mz)^2+(lz/t)^2}+{{(mz)^2+(lz/t)^2}\over 3}
\right)\e^{-\sqrt{(mz)^2+(lz/t)^2}}\nonumber \\
&&\times
\cos[m(\theta_i-\alpha)]
\cos(l(\varphi_i+2\pi\eta))
\Biggr],
\label{shiki59}
\end{eqnarray}
We choose the gauge group $SU(2)$ and consider the parameter
region of $t\sim z\sim 1$, which is the most interesting one
because the effects of both the temperature and the scale of the
extra dimension equally contribute to the effective potential.
As an illustration, the behavior of the total effective potential 
for $(t,~z)=(1.1,~0.8), \alpha=0$ with the $SU(2)$ fundamental
fermion is depicted in Fig.$6$. 
\begin{figure}[ht]
\begin{center}
\includegraphics[width=9cm,height=8cm,keepaspectratio]
{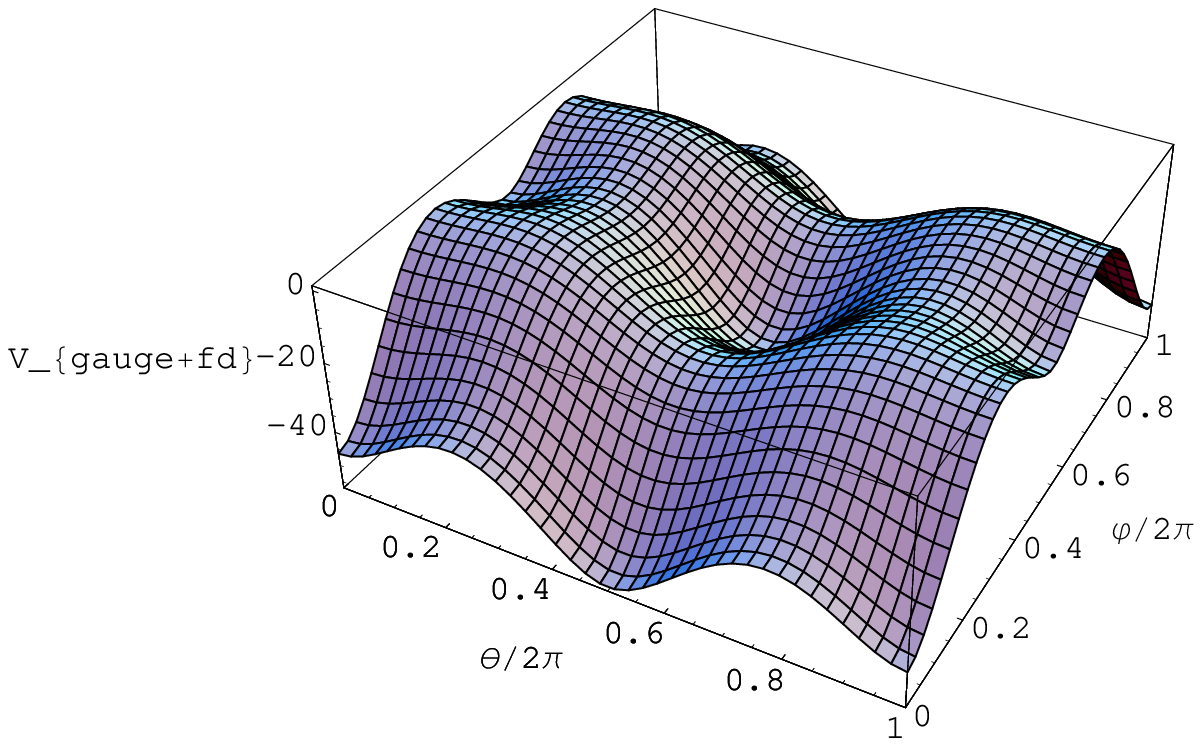}
\end{center}
\caption{The behavior of $N_{gauge}{\bar F}_{gauge}
+4N_{fd}{\bar F}_{fd}$ for $(N_{gauge},N_{fd})=(3,1), 
\alpha=0, LT=1.1, LM=0.8$. The minimum 
of the potential is given by $(\varphi,~\theta)
=(0, 0.5)\times 2\pi~({\rm mod}~2\pi)$}
\label{fig6}
\end{figure}
We find that the vacuum configuration is given by
\begin{equation}
(\varphi,~\theta)=(0,~\pi)~~({\rm mod}~~2\pi).
\label{shiki60}
\end{equation}
If we vary the parameter $\alpha$, the vacuum configuration 
changes according to the value, and we find that 
\begin{equation}
(\varphi,~\theta)=
\left\{\begin{array}{lll}
(0,~\pi) & \mbox{for}  & 0\leq \alpha< {\pi\over 2},\\[0.3cm]
%
%
(0,~0) & \mbox{for} & {\pi\over 2}\leq  \alpha \leq \pi.
\end{array}\right.
\label{shiki61}
\end{equation}
As stated below Eq.(\ref{shiki56}), the periodicity with
respect to $\theta$ for the fundamental fermion contribution 
becomes $\pi$ at $\alpha=\pi/2$. Again, nontrivial values of $\varphi$
are not realized and the $SU(2)$ gauge symmetry is not broken in this 
case. 
\par
Let us study the case of the massive adjoint fermion coupled to the
gauge field.  For $(t,~z)=(1.1,~0.8), \alpha=0$, we 
numerically find that the vacuum configuration is given by
\begin{equation} 
\left(\varphi,~\theta\right)
=\left(0,~0\right),~(0,~\pi)\quad({\rm mod}~~2\pi).
\label{shiki62}
\end{equation}
We next vary the parameters $\alpha$ and $N_{adj}$. 
The vacuum configuration $(\varphi,~\theta)=(0,~0)$ 
does not depend on the values of $\alpha$ for 
$N_{adj}=1, 2$. It changes according to the values 
of $\alpha$ for $N_{adj}\geq 3$. For $N_{adj}=3$ we obtain that
the vacuum configuration is given by
\begin{equation}
(\varphi,~\theta)=
\left\{\begin{array}{lll}
(0,~\pi/2) & \mbox{for}  & 0 \leq \alpha<\alpha_c, \\[0.3cm]
(0,~0) & \mbox{for}      &   \alpha_c<\alpha \leq \pi,
\end{array}\right.
\label{shiki63}
\end{equation}
where $\alpha_c/2\pi\simeq 0.1336$. If we consider $N_{adj}=6$, the
critical value is $\alpha_c/2\pi\simeq 0.1956$. 
\par
We have also numerically calculated the effective potential for
the other values of $(t,~z)$ with $t\sim z\sim 1$. It turns out
that the qualitative features are essentially the same as those
for $(t,~z)=(1.1,~0.8)$.
\par
Even if we consider the massive adjoint and fundamental fermions
simultaneously, we do not have nontrivial values for $\varphi$. The
VEV for $\theta$ depends on the size of the bulk mass and the flavor
number introduced in the theory. 
%
%
The gauge boson becomes massive only through the VEV of $\theta$ and the
$SU(2)$ gauge symmetry can be broken only by the Hosotani mechanism. 
\par
The boundary condition is crucial for determining the vacuum
configuration of $\varphi$ and $\theta$. 
The boundary condition for the $S_{\tau}^1$ direction is uniquely 
fixed by the quantum statistics and nontrivial 
VEVs for $\varphi$ cannot be realized, as we have studied
above. This is also valid for the $SU(N)$ gauge group 
with $N\geq 3$. On the other hand, the boundary condition for the
$S^1$ direction is controlled by the parameter $\alpha$, on which the VEV
for $\theta$ depends, in addition to the flavor number introduced into
the theory. 
\par
Let us make a comment. We have studied the 
VEVs of $\varphi_i$ and $\theta_i$ by minimizing the
effective potential. By using these values, we evaluate the 
second derivatives of the effective potential at the vacuum 
configuration \footnote{The off diagonal 
element $\del^2 V_{eff}/\del\theta_i \del\varphi_j$ vanishes for the vacuum
configuration.},
\begin{equation}
{g^2\over T^2}{\del^2 V_{eff}\over
 \del\varphi_i\del\varphi_j}\Bigg|_{vac},
\quad
{g^2L^2}{\del^2 V_{eff}\over
 \del\theta_i\del\theta_j}\Bigg|_{vac}.
\label{shiki64}
\end{equation}
These give us the gauge invariant (with respect to the 
residual gauge symmetry) mass terms for the zero modes 
for $A_{\tau}, A_y$, respectively. The mass term for the zero mode 
for $A_{\tau}$ is the electric mass and the one for $A_y$ is the 
scalar mass.
For instance, the electric mass from the effective 
potential (\ref{shiki10}) and (\ref{shiki14}) for $4$ dimensions is
calculated as
\begin{equation}
m_{ele}^2\equiv {g^2\over T^2}
{\del^2\over \del\varphi_i\del\varphi_j}
(V_{gauge}^T+V_{fd}^T)\Bigg|_{vac}
={g^2\over 3}T^2\left(2N+N_{fd}\right)M_{ij},
\label{shiki65}
\end{equation}
where the vacuum configuration in this case is given by
Eq.(\ref{shiki22}). Here $N_{fd}$ is the number of the massless
fundamental fermions and all the (off-) diagonal 
elements in the $N-1$ by $N-1$ matrix 
$M_{ij}$ is 2 (1). If we rescale the variables
$\varphi_i~(i=1,\cdots N-1), \varphi_N$ as
$\varphi_i \rightarrow \varphi_i/\sqrt{2}, 
\varphi_N\rightarrow \varphi_N/\sqrt{2N}$, the eigenvalue for the
matrix is given by $1/2$ ($(N-1)$-degeneracy). Therefore, the electric
mass is ${g^2\over 3}T^2\left(N+N_{fd}/2\right)$, which is the same
result obtained in \cite{gross}.
\par
Before closing this section, it is worthwhile mentioning the
high temperature behavior of the Hosotani mechanism.
At high temperature, broken symmetries via the Higgs mechanism
are expected to be restored\footnote{
For some special cases, the inverse symmetry breaking can
occur, as shown in the second reference \cite{finitet}.
}
because positive temperature-dependent mass squared terms
are induced radiatively \cite{finitet}.
This is also true for the time component of the gauge
field $A_{\tau}$, as shown in Eq. (\ref{shiki65}).
This does not, however, hold for the extra dimensional
component of the gauge field $A_y$.
The curvature at the origin of the potential is given as
\begin{equation}
{g^2L^2}{\del^2 V_{eff}\over
 \del\theta_i\del\theta_j}\Bigg|_{\theta=\varphi=0}.
\label{massA_y}
\end{equation}
To find the curvature, let us consider the effective
potential (\ref{shiki47}).
In the high temperature limit, the second term in Eq.(\ref{shiki47}) 
dominates and gives $\varphi=0$ as the vacuum configuration (we
have ignored the color indices here for simplicity.). 
For $\varphi=0$, the third terms for the fermion ($f=1, \eta=1/2$)
and for the boson ($f=0, \eta=0$) become
\begin{eqnarray}
2(LT)^D\sum_{l=1}^{\infty}
{(-1)^l\cos[m(\theta_i-\alpha)]\over [l^2 + (mLT)^2]^{D/2}}
&=&
-{\cos[m(\theta_i-\alpha)]\over m^D}
+(LT)^D\sum_{k=-\infty}^{\infty}
{{2\sqrt{\pi}}\over \Gamma(D/2)}
\Biggl[\biggl({{(k+\half)\pi}\over {mLT}}
\biggr)^2\Biggr]^{D-1\over 4}\nonumber\\
&&\times\ K_{{D-1}\over 2}\Bigl((2k+1)\pi mLT\Bigr)
\cos[m(\theta_i-\alpha)],
\label{newshikif}\\[0.8cm]
-2(LT)^D\sum_{l=1}^{\infty}
{\cos[m(\theta_i-\alpha)]\over [l^2 + (mLT)^2]^{D/2}}
&=&
{\cos[m(\theta_i-\alpha)]\over m^D}-(LT)^D
\Biggl({{\Gamma({D-1\over 2})\sqrt{\pi}}\over \Gamma(D/2)}
{1\over (mLT)^{D-1}}\nonumber\\
&+&{{2\sqrt{\pi}}\over \Gamma(D/2)}
\sum_{k=-\infty,\neq 0}^{\infty}
\Biggl[\biggl({\pi k\over mLT}\biggr)^2\Biggr]^{D-1\over 4}
K_{{D-1}\over 2}\Bigl(2k\pi mLT\Bigr)\Biggr)\nonumber\\
&&\times\,\cos[m(\theta_i-\alpha)],
\label{newshikib}
\end{eqnarray}
respectively. The first terms in 
Eqs.(\ref{newshikif}), (\ref{newshikib}) are 
canceled by the first term in Eq.(\ref{shiki47}). Since the 
modified Bessel function
$K_{{D-1}\over 2}(z)$ is exponentially suppressed for 
large $z$, we observe from Eq.(\ref{newshikif}) that the 
fermions do not contribute to the dynamics of $\theta$
at high temperature. 
\begin{equation}
F_{fermion}(\theta_i,\varphi=0) \simeq 0.
\label{F_fermion}
\end{equation}
Here, we have ignored the $\theta$-independent terms.
On the 
other hand, the second term in Eq.(\ref{newshikib}) survives to 
control the dynamics of $\theta$. Let us note that the 
difference between the
behavior in Eq.(\ref{newshikif}) and Eq.(\ref{newshikib}) comes
from the non-existence and the existence of the zero mode 
in the Matsubara frequency, respectively. 
We obtain from Eq.(\ref{newshikib}) that the bosonic 
contribution to the effective potential at high temperature
is given by
\begin{equation}
F_{boson}(\theta_i,\varphi=0) \simeq -{T\over L^{D-1}}
{\Gamma({D-1\over 2})\over \pi^{{D-1\over 2}}}\sum_{m=1}^{\infty}
{1\over m^{D-1}}\cos[m(\theta_i-\alpha)].
\label{newshiki2}
\end{equation}
Therefore, we have quite interesting results that there is
no fermionic contribution to the curvature (\ref{massA_y}) at 
high temperature
and that the bosonic contribution to the curvature is proportional
to $T$, but the coefficients can be both positive and negative
due to the boundary condition parametrized by $\alpha$ 
in Eq.(\ref{newshiki2}). This implies that broken 
symmetries via the Hosotani mechanism
at $T=0$ is not necessarily restored at high temperature, unlike
the Higgs mechanism.
Hence, we understand that, at high temperature, only the bosonic
degrees of freedom determine 
the gauge symmetry breaking patterns, 
which depend on the bosonic
matter contents and the boundary 
conditions for the $S^1$ direction \cite{sakatake}. 
\par
\section{Conclusions}
We have studied the gauge theories with/without the extra
dimension at finite temperature, and especially focused on 
the zero mode of the component gauge field $A_{\tau}$ for the
Euclidean time direction. The zero
mode is closely related with the Polyakov loop, and 
we have computed the effective potential for the
zero mode in the one-loop approximation. We minimize the effective 
potential to study whether nontrivial values 
for $\vev{A_{\tau}}$ are realized or not.
\par
The vacuum structure crucially depends on 
the boundary conditions of the fields for the
compactified direction. In the present case, the boundary
condition of the field for the Euclidean 
time direction is uniquely fixed by the quantum statistics.
This is a big difference from the case of the boundary condition 
of the field for the spatial compact extra dimension. 
In the pure $SU(N)$ gauge theory and the $SU(N)$ gauge theory with
the massless adjoint matter, the Polyakov loop takes 
the values at the center of the $SU(N)$, and this is
consistent with the lattice result in the high temperature region for $SU(3)$. 
For the fundamental massless matter coupled to the gauge field, no 
nontrivial values for $\vev{A_{\tau}}$ are induced, so that the gauge
symmetry is not broken and the gauge bosons remain massless.
The boundary condition for the Euclidean time direction 
prevents $\vev{A_{\tau}}$ from taking nontrivial values.
\par
We have also considered the massive bulk matter to see the effect
of the bulk mass on $\vev{A_{\tau}}$. The matter with $M/T \gg 1$
decouples from the effective potential due to the Boltzmann
factor. Although a small bulk mass tends to induce nontrivial
VEVs, the effect is too small to realize the gauge symmetry
breaking.
\par
In order to investigate further the possibility of having nontrivial
VEVs for $A_{\tau}$, we have considered one spatial extra dimension at
finite temperature, which is compactified on $S^1$, and have 
studied the gauge theories on $S^1_{\tau}\times R^3\times S^1$. There 
are two kinds of the order parameters in this case, that is, $\theta_i$ 
and $\varphi_i$, as given in Eq.(\ref{shiki40}). The Wilson 
loop and the Polyakov loop are the relevant quantities 
for the dynamics. We have computed the effective potential
for the order parameters along the flat direction (\ref{shiki39}) and 
minimize it to determine the vacuum configuration. The boundary conditions 
for the $S^1$ direction is parametrized by $\alpha$, and those for
the $S_{\tau}^1$ direction is uniquely fixed by the quantum
statistics. The effective potential (\ref{shiki42}) is regarded
as a special case of the six dimensional gauge theory
compactified on $T^2$ with appropriate boundary 
conditions. As far as our numerical
analyses are concerned, no nontrivial values 
for $\vev{A_{\tau}}$ are realized and the gauge symmetry 
breaking can occur only through nontrivial values for 
$\theta_i$. 
%
%
\par
In our analyses, the gauge bosons become massive only through
$\theta_i$; that is, the gauge symmetry is broken by the 
Hosotani mechanism. We do not find the models in which the 
gauge symmetry is broken through the VEV 
of $A_{\tau}$. No nontrivial values of $\vev{A_{\tau}}$ are
obtained. As long as the boundary condition for the Euclidean time 
direction is fixed by the quantum statistics, our analyses
strongly suggest that it is impossible to break dynamically the
gauge symmetry through $\vev{A_{\tau}}$ in perturbation theory.
It may be challenging to find models in which the 
Polyakov loop $W_p$ in Eq.(\ref{shiki12}) takes
nontrivial values nonperturbatively. 
\begin{center}
{\bf Acknowledgments}
\end{center}
This work is supported in part by a Grant-in-Aid for Scientific Research
(No. 18540275) from the Japanese Ministry of Education, Science, Sports
and Culture. The authors would like to thank 
Professors T. Onogi(Yukawa Inst.), H. So(Ehime Univ.) and
H. Yoneyama(Saga Univ.) for valuable discussions and K.T. is 
also supported by the 21st Century COE 
Program at Tohoku University.
\vspace*{1cm}

\end{document}